\RequirePackage[mathlines]{lineno}
\documentclass[twocolumn,showpacs,preprintnumbers,amsmath,amssymb,aps,prc,superscriptaddress,nofootinbib]{revtex4}

\usepackage{graphicx}
\usepackage{dcolumn}
\usepackage{bm}
\usepackage{color}

\newcommand{\sqrtsNN} {\mbox{$\sqrt{\mathrm{\it s_{NN}}}$}}

\newcommand{\cor} {\mbox{$\langle \cos [3(\phi_j-\phi_i)] \rangle_{i \neq j} $}}
\newcommand{\corpT} {\mbox{$\langle \cos [3(\phi_j(p_T)-\phi_i)]\rangle_{i \neq j} $}}
\def \GeVc {\mbox{$\mathrm{GeV} / c$}}
\def \lt {\mbox{$<$}}
\def \gt {\mbox{$>$}}
\def \etal {\mbox{$\mathrm{\it et\ al.}$}}
\def \deta {\mbox{$\Delta \eta$ }}
\def \deleta {\mbox{$\Delta \eta$}}

\newcommand{ \mean }[1]{\left\langle #1 \right\rangle}

\begin{document}

\title{Third Harmonic Flow of Charged Particles in Au+Au Collisions at $\sqrtsNN = 200$ GeV}

%
\affiliation{AGH University of Science and Technology, Cracow, Poland}
\affiliation{Argonne National Laboratory, Argonne, Illinois 60439, USA}
\affiliation{University of Birmingham, Birmingham, United Kingdom}
\affiliation{Brookhaven National Laboratory, Upton, New York 11973, USA}
\affiliation{University of California, Berkeley, California 94720, USA}
\affiliation{University of California, Davis, California 95616, USA}
\affiliation{University of California, Los Angeles, California 90095, USA}
\affiliation{Universidade Estadual de Campinas, Sao Paulo, Brazil}
\affiliation{Central China Normal University (HZNU), Wuhan 430079, China}
\affiliation{University of Illinois at Chicago, Chicago, Illinois 60607, USA}
\affiliation{Cracow University of Technology, Cracow, Poland}
\affiliation{Creighton University, Omaha, Nebraska 68178, USA}
\affiliation{Czech Technical University in Prague, FNSPE, Prague, 115 19, Czech Republic}
\affiliation{Nuclear Physics Institute AS CR, 250 68 \v{R}e\v{z}/Prague, Czech Republic}
\affiliation{University of Frankfurt, Frankfurt, Germany}
\affiliation{Institute of Physics, Bhubaneswar 751005, India}
\affiliation{Indian Institute of Technology, Mumbai, India}
\affiliation{Indiana University, Bloomington, Indiana 47408, USA}
\affiliation{Alikhanov Institute for Theoretical and Experimental Physics, Moscow, Russia}
\affiliation{University of Jammu, Jammu 180001, India}
\affiliation{Joint Institute for Nuclear Research, Dubna, 141 980, Russia}
\affiliation{Kent State University, Kent, Ohio 44242, USA}
\affiliation{University of Kentucky, Lexington, Kentucky, 40506-0055, USA}
\affiliation{Institute of Modern Physics, Lanzhou, China}
\affiliation{Lawrence Berkeley National Laboratory, Berkeley, California 94720, USA}
\affiliation{Massachusetts Institute of Technology, Cambridge, MA 02139-4307, USA}
\affiliation{Max-Planck-Institut f\"ur Physik, Munich, Germany}
\affiliation{Michigan State University, East Lansing, Michigan 48824, USA}
\affiliation{Moscow Engineering Physics Institute, Moscow Russia}
\affiliation{National Institute of Science Education and Research, Bhubaneswar 751005, India}
\affiliation{Ohio State University, Columbus, Ohio 43210, USA}
\affiliation{Old Dominion University, Norfolk, VA, 23529, USA}
\affiliation{Institute of Nuclear Physics PAN, Cracow, Poland}
\affiliation{Panjab University, Chandigarh 160014, India}
\affiliation{Pennsylvania State University, University Park, Pennsylvania 16802, USA}
\affiliation{Institute of High Energy Physics, Protvino, Russia}
\affiliation{Purdue University, West Lafayette, Indiana 47907, USA}
\affiliation{Pusan National University, Pusan, Republic of Korea}
\affiliation{University of Rajasthan, Jaipur 302004, India}
\affiliation{Rice University, Houston, Texas 77251, USA}
\affiliation{Universidade de Sao Paulo, Sao Paulo, Brazil}
\affiliation{University of Science \& Technology of China, Hefei 230026, China}
\affiliation{Shandong University, Jinan, Shandong 250100, China}
\affiliation{Shanghai Institute of Applied Physics, Shanghai 201800, China}
\affiliation{SUBATECH, Nantes, France}
\affiliation{Temple University, Philadelphia, Pennsylvania, 19122, USA}
\affiliation{Texas A\&M University, College Station, Texas 77843, USA}
\affiliation{University of Texas, Austin, Texas 78712, USA}
\affiliation{University of Houston, Houston, TX, 77204, USA}
\affiliation{Tsinghua University, Beijing 100084, China}
\affiliation{United States Naval Academy, Annapolis, MD 21402, USA}
\affiliation{Valparaiso University, Valparaiso, Indiana 46383, USA}
\affiliation{Variable Energy Cyclotron Centre, Kolkata 700064, India}
\affiliation{Warsaw University of Technology, Warsaw, Poland}
\affiliation{University of Washington, Seattle, Washington 98195, USA}
\affiliation{Wayne State University, Detroit, Michigan 48201, USA}
\affiliation{Yale University, New Haven, Connecticut 06520, USA}
\affiliation{University of Zagreb, Zagreb, HR-10002, Croatia}

\author{L.~Adamczyk}\affiliation{AGH University of Science and Technology, Cracow, Poland}
\author{J.~K.~Adkins}\affiliation{University of Kentucky, Lexington, Kentucky, 40506-0055, USA}
\author{G.~Agakishiev}\affiliation{Joint Institute for Nuclear Research, Dubna, 141 980, Russia}
\author{M.~M.~Aggarwal}\affiliation{Panjab University, Chandigarh 160014, India}
\author{Z.~Ahammed}\affiliation{Variable Energy Cyclotron Centre, Kolkata 700064, India}
\author{A.~V.~Alakhverdyants}\affiliation{Joint Institute for Nuclear Research, Dubna, 141 980, Russia}
\author{I.~Alekseev}\affiliation{Alikhanov Institute for Theoretical and Experimental Physics, Moscow, Russia}
\author{J.~Alford}\affiliation{Kent State University, Kent, Ohio 44242, USA}
\author{C.~D.~Anson}\affiliation{Ohio State University, Columbus, Ohio 43210, USA}
\author{D.~Arkhipkin}\affiliation{Brookhaven National Laboratory, Upton, New York 11973, USA}
% ...
\collaboration{STAR Collaboration}\noaffiliation

\begin{abstract}
We report measurements of the third harmonic coefficient of the azimuthal anisotropy, $v_3$, known as triangular flow. The analysis is for charged particles in Au+Au collisions at $\sqrtsNN = 200$ GeV, based on data from the STAR experiment at the BNL Relativistic Heavy Ion Collider. Two-particle correlations as a function of their pseudorapidity separation are fit with narrow and wide Gaussians. Measurements of triangular flow are extracted from the wide Gaussian, from two-particle cumulants with a pseudorapidity gap, and also from event plane analysis methods with a large pseudorapidity gap between the particles and the event plane. These results are reported as a function of transverse momentum and centrality. A large dependence on the pseudorapidity gap is found. Results are compared with other experiments and model calculations.
\end{abstract}


\maketitle

\section{Introduction}
The study of azimuthal anisotropy, based on Fourier coefficients, is recognized as an important tool to probe the hot, dense matter created in heavy-ion collisions~\cite{Voloshin:2008dg,Sorensen:2009cz}.  The first harmonic coefficient $v_1$, called directed flow, and the second harmonic coefficient $v_2$, called elliptic flow, have been extensively studied both experimentally and theoretically, while higher even-order harmonics have also garnered some attention~\cite{v4v6}.  In contrast, odd harmonics of order three and above were overlooked until recently~\cite{Mishra:2007tw, geoFluct1}. This is because in a picture with smooth initial overlap geometry, it had been assumed that higher-order odd harmonics are required to be zero by symmetry. More recently it has been realized that event-by-event fluctuations break this symmetry~\cite{geoFluct1, derik, Sorensen:2011hm}. The event plane of the detected particles approximates the plane of the participating particles and for reasonable event-plane resolutions the measured $v_n$ are not the mean values, but closer to the root-mean-square values~\cite{Ollitrault:2009ie}. As a consequence, higher-order odd harmonics carry valuable information about ``hot spots" or ``lumpiness" in the initial state of the colliding system~\cite{Mishra, geoFluct2, transpov3, v3_4D-hydro, hydrov3, Schenke:2011bn, v3-AMPT, Schenke:2012wb, Gale:2012rq}.

The third harmonic coefficient -- sometimes called triangular flow, but probably not related to triangular configurations in the initial state -- is thus a new tool to study initial state fluctuations and the subsequent evolution of the collision system. It is probably related to the production of the near-side ridge~\cite{geoFluct1,Voloshin:2011mx} observed when correlations are studied as a function of the difference of azimuthal angles and the difference of pseudorapidities of the particles. Theoretical studies suggest that $v_3$ is more sensitive to viscous effects than $v_2$ because the finer details of the higher harmonics are smoothed more by viscosity~\cite{transpov3}. It also appears that the mean value of the initial state triangular eccentricity in coordinate space, from central to midcentral collisions, is independent of the geometric model used for the initial overlap~\cite{Bhalerao:2011yg}, unlike the second harmonic spatial elliptic eccentricity. This is probably because $v_3$ is an odd harmonic and dominated by fluctuations. Rapidity-even $v_1$ is symmetric about midrapidity and is also dominated by fluctuations, but is complicated by the correction needed for conservation of momentum~\cite{Luzum:2010fb}. Higher odd harmonics are thought to be less useful because of non-linear terms coming from the eccentricities of lower harmonics~\cite{Teaney:2012ke}. Thus $v_3$ is an ideal flow harmonic to study viscosity because it is almost insensitive to the model used for the initial conditions and more sensitive to viscosity.

In order to separate the long-range correlations of interest from short-range correlations, we present measurements, based on the azimuthal angle $\phi$, of $\cor$ vs.\ the pseudorapidity separation \deta = $\eta_{i} - \eta_{j}$ between the two particles (i, j), fit with narrow and wide Gaussians. We present results derived from the wide Gaussian, for two-particle cumulants~\cite{Cumulant}, and for the standard event plane methods~\cite{methodPaper}, as a function of transverse momentum $p_{T}$, pseudorapidity gap \deta, and centrality. The pseudorapidity gap between the particles being correlated is found to be an especially important experimental variable. We compare our results to other experiments, and to both transport and hydrodynamic models.

\begin{figure}
\resizebox{0.5\textwidth}{!}{
  \includegraphics{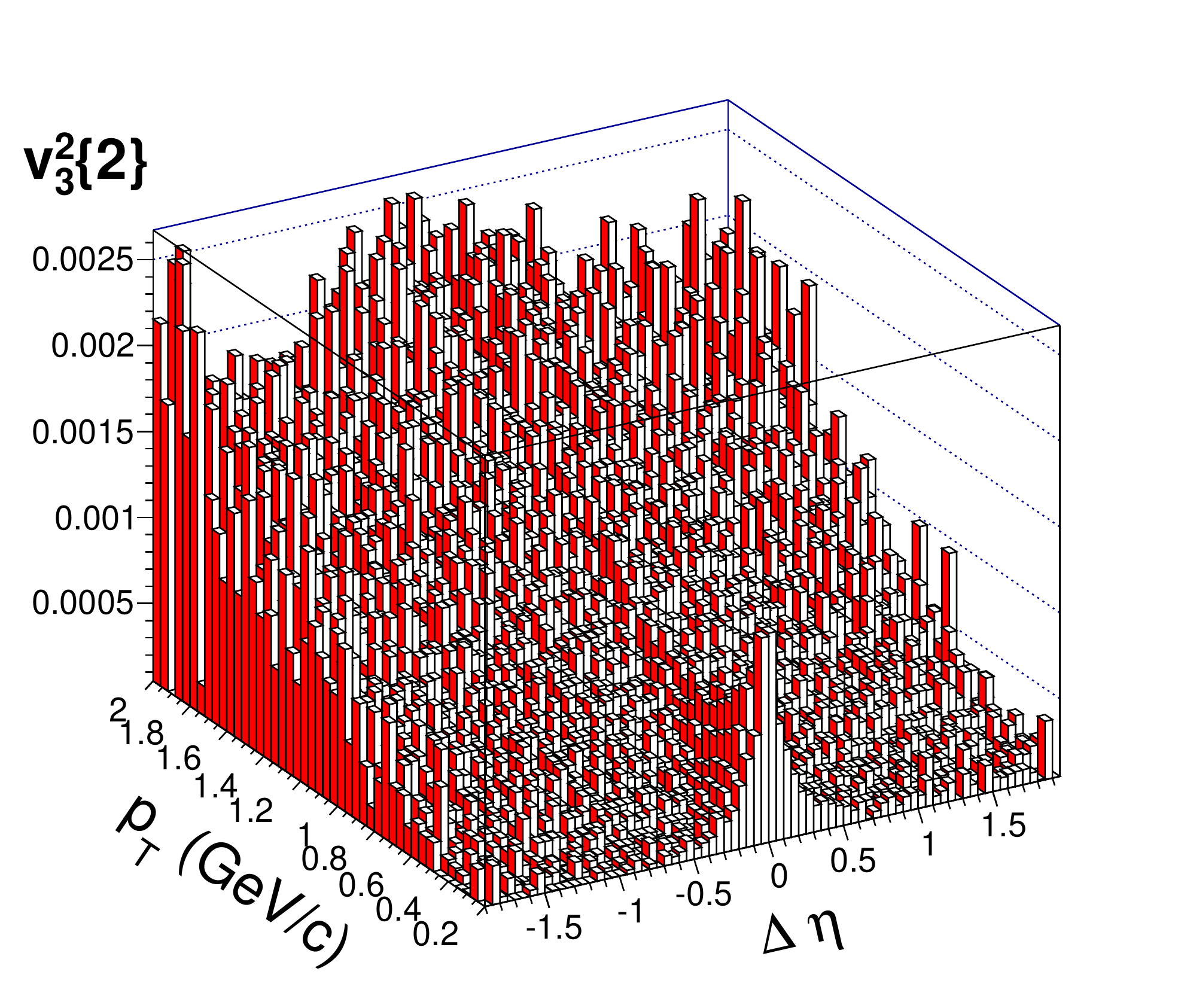} }
\caption {(color online) $v_{3}^{2}\{2\}$ vs.\ transverse momentum and pseudorapidity separation for charged hadron pairs in 200 GeV Au+Au minimum bias collisions. The $p_T$ is for one of the particles, integrated over the $p_T$ values of the other particle in the range $0.15 \lt p_T \lt 2.0$ \GeVc.}
\label{fig:2D}      
\end{figure}

\section{Experiment}
About ten million Au+Au collisions at $\sqrtsNN = 200$ GeV have been used in this study, all acquired in the year 2004 using the STAR detector with a minimum bias trigger.  The main Time Projection Chamber (TPC)~\cite{TPC} of STAR covers pseudorapidity $|\eta| \lt 1.0$, while two Forward Time Projection Chambers (FTPCs)~\cite{FTPC} cover $2.5 \lt |\eta| \lt 4.0$. The extended range in $\eta$ of the FTPCs was important because the analyses were done as a function of the $\eta$ gap between particles. This requirement limited the study to the data collection years when the FTPCs were operational. The centrality definition of an event is based on the number of charged tracks in the TPC with track quality cuts of $|\eta| \lt 0.5$, a distance of closest approach (DCA) to the primary vertex less than 3 cm, and 15 or more space points out of a total of 45. This analysis used events with vertex $z$ coordinate (along the beam direction) within 30 cm from the center of the TPC. For each centrality bin, the number of participants and binary collisions can be found in Table III of Ref.~\cite{Agakishiev:2011eq}.

\section{Analysis Methods}

\subsection{Event Planes}
In the standard event plane method~\cite{methodPaper} for $v_3$, we reconstruct a third harmonic event plane $\Psi_{3}$ from TPC tracks and also from FTPC tracks.  For event plane reconstruction, we use tracks with transverse momentum $p_T  \gt  0.15$ GeV$/c$, that pass within 3 cm of the primary vertex, and have at least 15 space points in the TPC acceptance $(|\eta| \lt 1.0)$ or 5 space points in the FTPC acceptance ($2.5 \lt |\eta| \lt 4.0$). It is also required that the ratio of the number of actual space points to the maximum possible number of space points along each track's trajectory be greater than 0.52.  In event plane calculations, tracks have a weighting factor $w = p_T$ in units of GeV$/c$ for $p_T \lt 2$ GeV$/c$, and $w = 2$ GeV$/c$ for $p_T \geq$ 2 GeV$/c$. Although the STAR detector has good azimuthal symmetry, small acceptance effects in the calculation of the event plane azimuth were removed by the method of shifting~\cite{Voloshin:2008dg}. When using the TPC event plane, we used the $\eta$ subevent method which provides an $\eta$ gap, but with an additional small $\eta$ gap of $\pm$0.05 between the subevents~\cite{methodPaper}. The $\eta$ subevent method avoids self-correlations because the particles and the event plane are in opposite hemispheres. When using the FTPCs, we obtained the subevent plane resolution from the correlation of the two FTPCs, but then used the full event plane from both FTPCs~\cite{methodPaper}. This introduced a large $\eta$ gap between the particles in the TPC and the FTPC event planes. Since there is no overlap between the coverage of the TPC and FTPCs, there is no possibility of self-correlation when using the FTPC event plane.

\subsection{2-particle correlations}

\begin{figure}
\resizebox{0.4\textwidth}{!}{
  \includegraphics{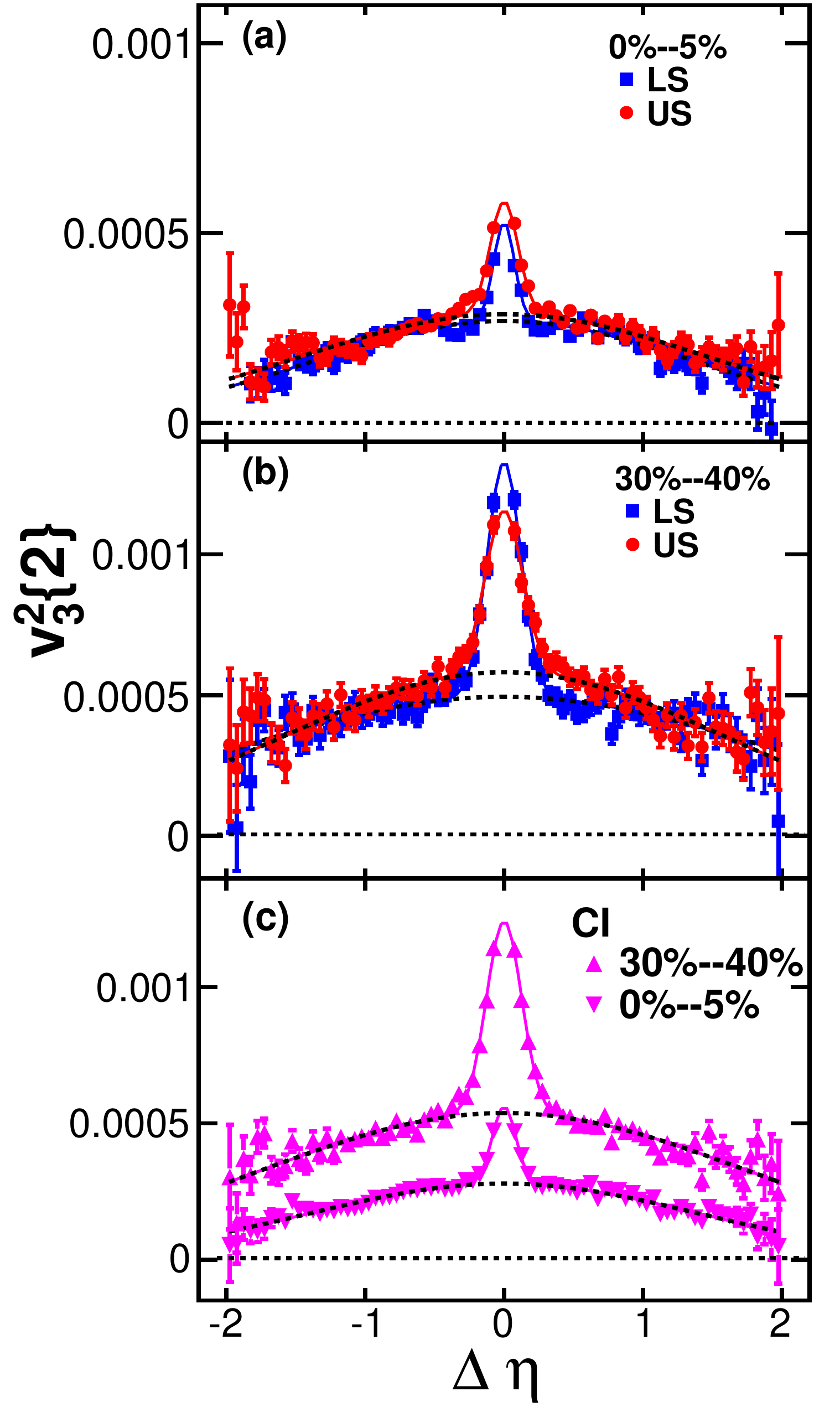} }
\caption {(color online) $v_{3}^{2}\{2\}$ vs.\ the pseudorapidity separation of the particles in pairs for charged hadrons with $0.15 \lt p_T \lt 2.0$ \GeVc\  within two centrality intervals in 200 GeV Au+Au collisions. Data are fit with narrow and wide Gaussians. Like Sign (LS), Unlike Sign (US), and Charge Independent (CI) cases are shown with only statistical errors. The dashed curves under the peaks are the wide Gaussian fits.}
\label{fig:gaus}      
\end{figure}
 
We studied $v_{3}^{2}\{2\}$  = $\cor$ vs.\ \deta  between the two particles. For this two-particle cumulant method~\cite{Cumulant}, acceptance correction terms, which were generally small, were evaluated and applied. Figure \ref{fig:2D} shows that there is a sharp peak for tracks close in $\eta$ and at low $p_T$. This has also been seen by PHOBOS~\cite{Alver:2010rt}. Our distribution of $v_{3}^{2}\{2\}$  vs.\ \deta can be well described by wide and narrow Gaussian peaks as shown in Fig.~\ref{fig:gaus} for two centrality intervals. Using two Gaussians plus a flat background gave the same results for $v_3$ when integrated for all accepted pairs within the range $|\Delta\eta|\lt2$, as described below. The narrow Gaussian is identified as short range nonflow correlations like the Bose-Einstein correlation, resonance decay, and Coulomb interactions, reduced by effects from track merging. The narrow peak disappears above $p_{T} \gt 0.8$ GeV/$c$, so is unlikely to be from jet correlations. The wide Gaussian is the signal of interest in this paper and its fit parameters are used to calculate $v_{3}^2\{2\}$ as a  function of centrality and transverse momentum for accepted pairs within the range $|\Delta\eta|\lt2$. The differential $v_{3}^{2}\{2\}$ can be averaged over $p_{T}$ and $\eta\lt1$ as,  

\begin{equation} 
  \left< v_{3}^{2}\{2\} \right>   =   \frac {\int^b_a v_{3}^{2}\{2\}  W d(\Delta \eta )} {\int^b_a W d(\Delta \eta)} ,
\label{eq:int} 
\end{equation}
where $W$ equals $dN/d(\Delta \eta) $ when weighted with the number of particle pairs. The integration ranges for numerator and denominator are the same. This is normally called the integrated $v_{3}^{2}\{2\}$. To evaluate the effect of weighting we also used unit weight $= 1$, which will be shown to make little difference. The differential $v_{3}\{2\}(p_T)$ can be obtained from the scalar product~\cite{Voloshin:2008dg} relation
\begin{equation} 
  v_{3}\{2\}(p_T)   =   \frac{\corpT} {\sqrt{\left< v_{3}^{2}\{2\} \right>}}  .
\label{eq:dif} 
\end{equation} 
where the $j^{\rm th}$ particle is selected from the $p_{T}$ bin of interest.
   
\begin{figure}
\resizebox{0.35\textwidth}{!}{
  \includegraphics{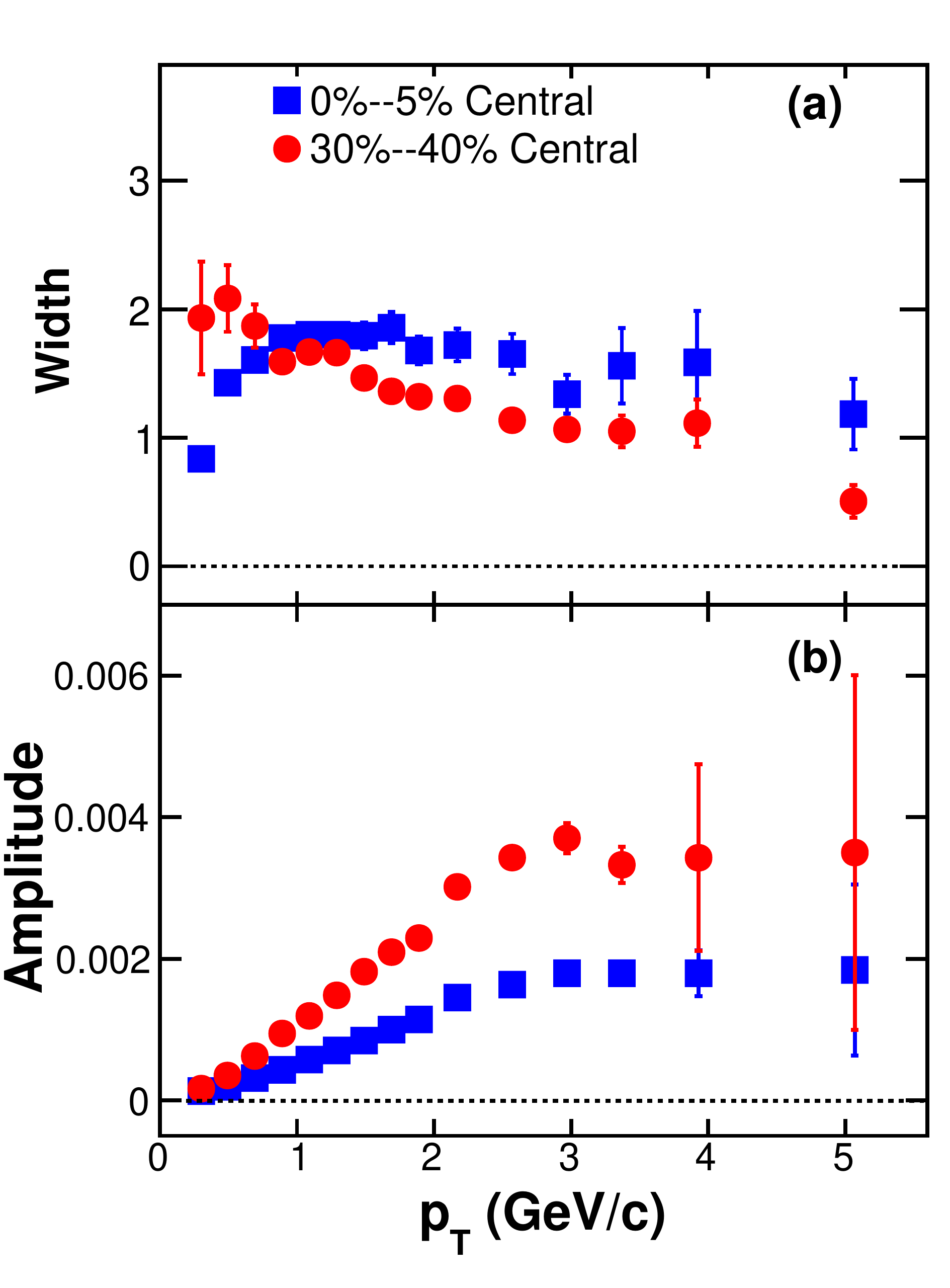} }
\caption{(color online) The width in units of $\Delta \eta$ and amplitude of the charge independent wide Gaussian as a function of transverse momentum for most central (0\%--5\%) and midcentral (30\%--40\%)  Au+Au collisions at $\sqrtsNN= 200$ GeV.  The plotted errors are statistical.}
\label{fig:pT_gaus}
\end{figure}

\begin{figure}
\resizebox{0.45\textwidth}{!}{
  \includegraphics{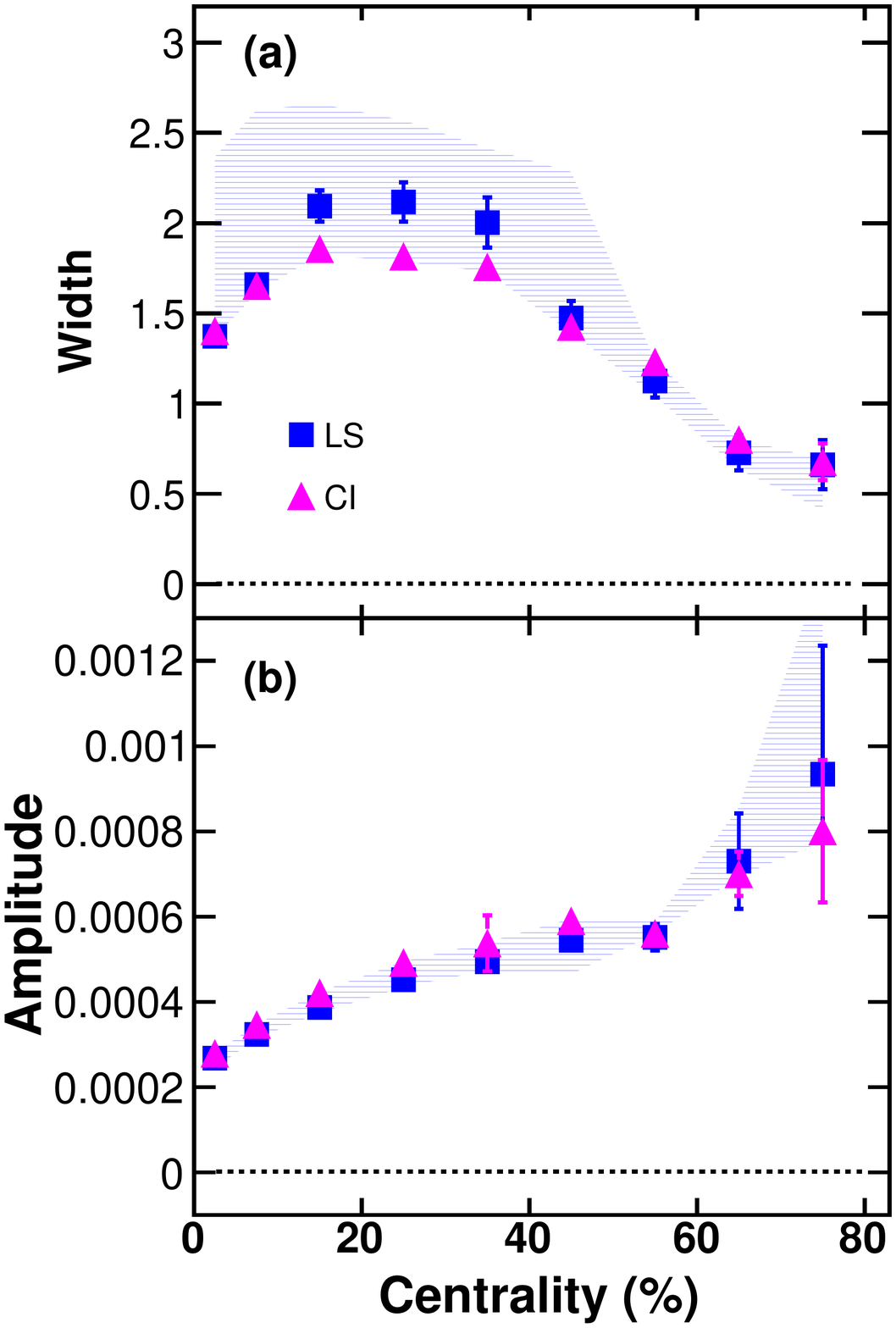} }
\caption{(color online) The width in units of $\Delta \eta$ and amplitude of the wide Gaussian as a function of centrality for Charge Independent (CI) and Like Sign (LS) particles with $0.15 \lt p_T \lt 2.0$ \GeVc\ for Au+Au collisions at $\sqrtsNN= 200$ GeV. The errors on the data points are statistical. The upper edge of the systematic error band for the Like Sign particles shows the width of the wide Gaussian required to also fit the data from the FTPC.}
\label{fig:cent_gaus}      
\end{figure}

Figure~\ref{fig:pT_gaus} shows the $p_T$ dependence of the width and amplitude of the wide Gaussian fit to the data in Fig.~\ref{fig:gaus}. Other functional forms, such as one with a constant offset are discussed below. Shown are results for the 0\%--5\% most central and 30\%--40\% midcentral collisions. Above 0.8 $\GeVc$ the distribution can be described by a single wide Gaussian. The amplitude increases with $p_T$ and then saturates around 3 $\GeVc$. The $p_{T}$ dependence of the width depends on centrality, with the 0\%--5\% most central data showing first an increase in the width and then a gradual decrease, while for the 30\%--40\% central data the width appears to gradually decrease for all $p_{T}$.

Figure~\ref{fig:cent_gaus} shows the centrality dependence of the width and amplitude of the wide Gaussian. In peripheral collisions, the Gaussian width is narrow and well constrained by the data. As the collisions become more central, the width broadens reaching beyond 1.5 units in pseudorapidity in the centrality range 10\%--40\%. When the width of the wide Gaussian becomes broader than $\Delta\eta=1$, it becomes difficult, with the data from the TPC alone, to distinguish between functional forms for $v_3\{2\}(\Delta\eta)$ with and without a background. The data points in Fig.~\ref{fig:cent_gaus} show the results when fitting a single wide Gaussian to the TPC data alone. The upper edge of the systematic error band corresponds to a width that would allow the fit function to extend out far enough to match the FTPC data at $\mean{\Delta\eta}=3.21$. On the other hand, if we include a constant background, we can also match the FTPC and TPC data with a wide Gaussian width consistent with the lower edge of the error band in Fig.~\ref{fig:cent_gaus}. A larger acceptance in $\eta$ is required to better constrain the functional form. Such a constraint could help distinguish between different physical mechanisms underlying the signal, such as stochastic fluctuations in the hydrodynamic phase~\cite{Kapusta:2011gt} or decoherence of flux-tube like structures in the longitudinal direction~\cite{Dusling}.

Whether using the TPC data only or also including the FTPC data, the width of the wide Gaussian peak tends to become more narrow for the most central collisions than is observed for midcentral collisions. The rise and then fall of the width of $v_3\{2\}(\Delta\eta)$ mimics the rise and fall of the low $p_T$ ridge amplitude reported in Ref.~\cite{Agakishiev:2011pe}. Reference~\cite{Sorensen:2011hm} describes this centrality trend in terms of participant eccentricity fluctuations, where the fluctuations in midcentral collisions are well above statistical expectations. This can be attributed to the asymmetry of the overlap region of the colliding nucleons which allows a nucleon on the periphery of one nucleus to impinge on many nucleons in the center of the other nucleus thus amplifying the effect of fluctuations of nucleon positions in the periphery of the nucleus. Thus the width of $v_3\{2\}(\Delta\eta)$ and the amplitude of the low $p_T$ ridge may be related to the same fluctuations.

\section{Results}

First we will show $v_3$ vs.\ $\eta$ using two standard event plane methods, followed by $v_3$ vs.\ $p_T$ for these methods and also for the wide Gaussian two-particle correlation. Finally, we present the integrated $v_3$ vs.\ centrality for these methods and also for the two-particle cumulant method~\cite{Cumulant} with an $\eta$ gap. Results in all the figures are presented with only statistical errors unless stated otherwise. 

\subsection{$\eta$ dependence}

\begin{figure}
\resizebox{0.5\textwidth}{!}{
  \includegraphics{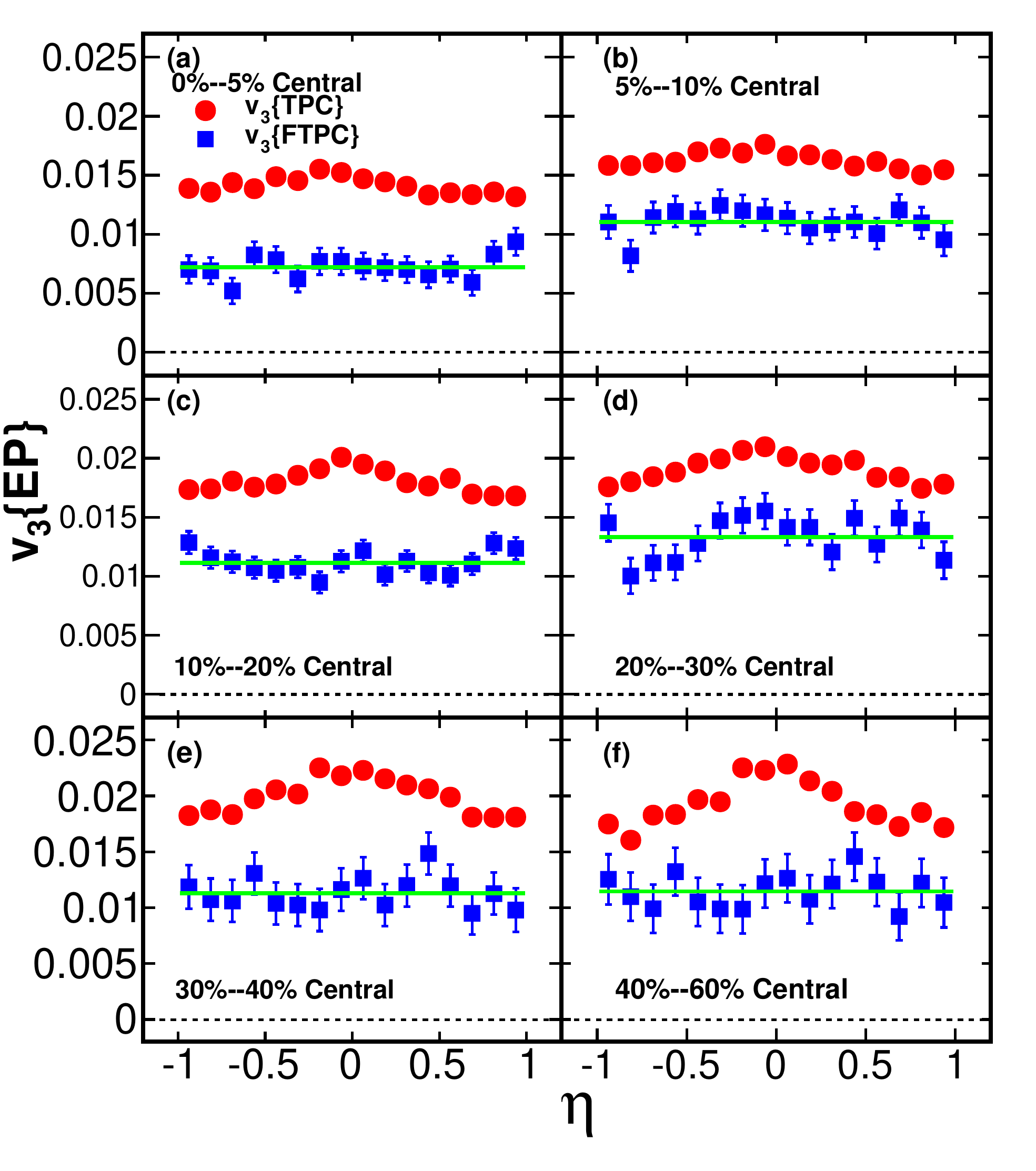} }
\caption {(color online) The third harmonic coefficient as a function of pseudorapidity for different centralities for Au+Au collisions at $\sqrtsNN= 200$ GeV, with track selection in the TPC of $0.15 \lt p_T \lt 2.0$ \GeVc. Results are shown for the event plane constructed either in the TPC or in the FTPCs. The horizontal lines are fits to the FTPC results.} 
\label{fig:eta}      
\end{figure}

Figure~\ref{fig:eta} shows the $\eta$ dependence of $v_3$ using two event plane methods. For particles in the TPC using the opposite $\eta$ subevent for the event plane, $v_3$ is slightly peaked at midrapidity. With the event plane in the FTPCs there is a large $\eta$ gap between the particles and the plane, and $v_3$ is flat for all centralities. This flatness means that acceptance effects at the edges of the TPC are not significant. Thus, even though a large \deta in Fig.~\ref{fig:gaus} means that one of the particles must be at large $\eta$ in Fig.~\ref{fig:eta}, this evidently is not a significant effect on the flatness of the \deta dependence.

\subsection{$p_T$ dependence}

\begin{figure}%
\resizebox{0.45\textwidth}{!}{
  \includegraphics{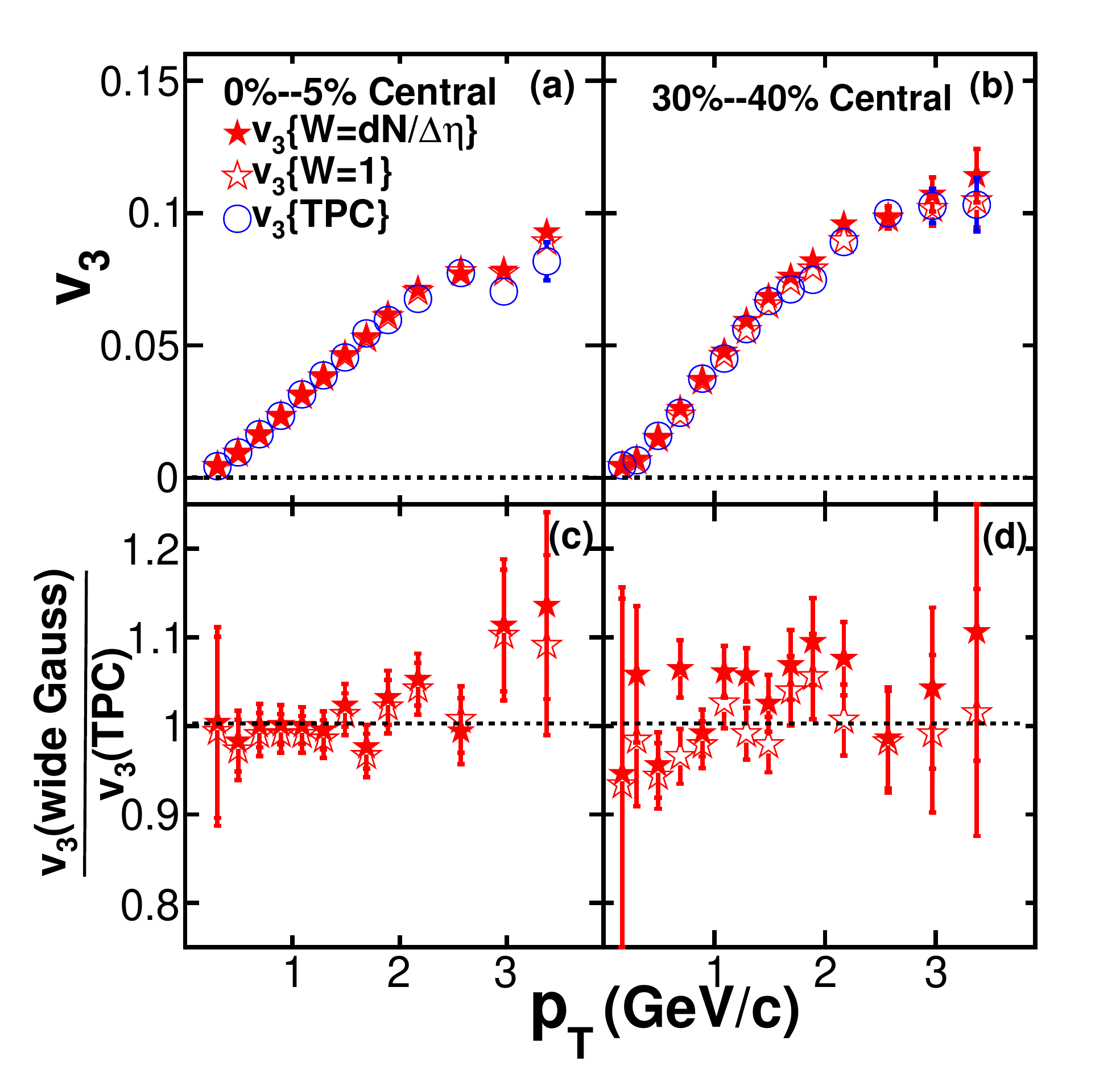} }
\caption {(color online) The top panels show third harmonic coefficient as a function of $p_T$ for the wide Gaussian method and for the event plane in the TPC, for two centralities for Au+Au collisions at $\sqrtsNN = 200$ GeV, for tracks in the TPC with $-1.0 \lt \eta \lt 1.0$. The wide Gaussian was weighted with either the number of particle pairs or by unity. The bottom panels show the ratio of $v_3$ from the wide Gaussian method to $v_3$ from the TPC subevent method.} 
\label{fig:pT}      
\end{figure}

\begin{figure}
\resizebox{0.47\textwidth}{!}{
  \includegraphics{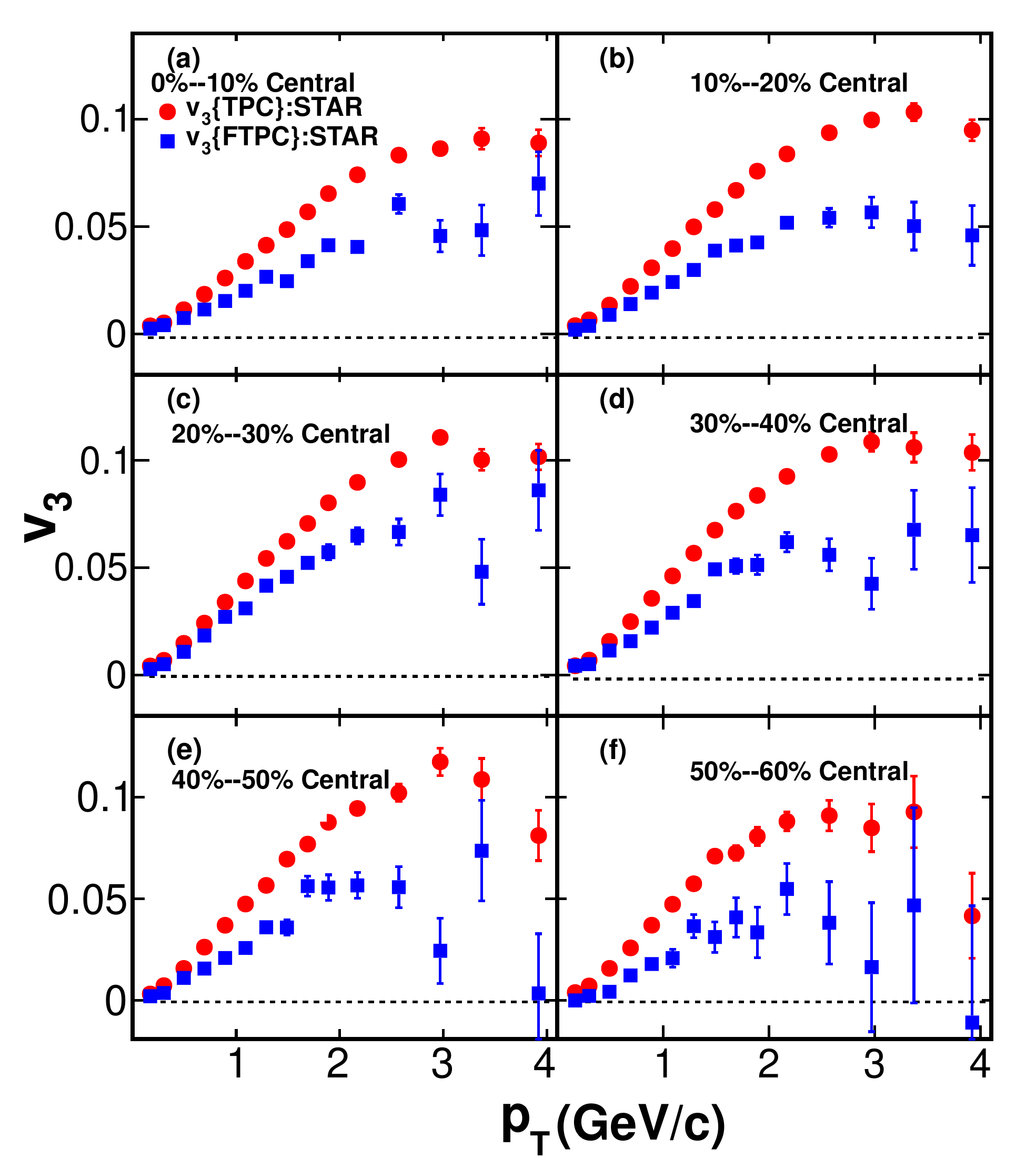} }
\caption {(color online) The third harmonic coefficient as a function of $p_T$, for different centralities for Au+Au collisions at $\sqrtsNN = 200$ GeV, for tracks in the TPC with $-1.0 \lt \eta \lt 1.0$. The event planes are constructed either in the TPC or in the FTPCs.}
\label{fig:FTPC}      
\end{figure}

The $p_T$ dependence is shown in Fig.~\ref{fig:pT}. For the wide Gaussian method, Eq.~(\ref{eq:dif}) was used together with the parameters from Fig.~\ref{fig:pT_gaus} for each $p_T$ bin. The results for the wide Gaussian method with either kind of weighting are almost the same as those for the TPC using subevent planes, meaning that for either of these two methods the narrow Gaussian does not significantly affect the wide Gaussian. However, in Fig.~\ref{fig:FTPC} the results with the event plane in the FTPCs are considerably lower, presumably because of the larger $\eta$ gap to be discussed in Sec.~\ref{sec:deltaEta}.

\subsection{Centrality dependence}

\begin{figure}
\resizebox{0.5\textwidth}{!}{
  \includegraphics{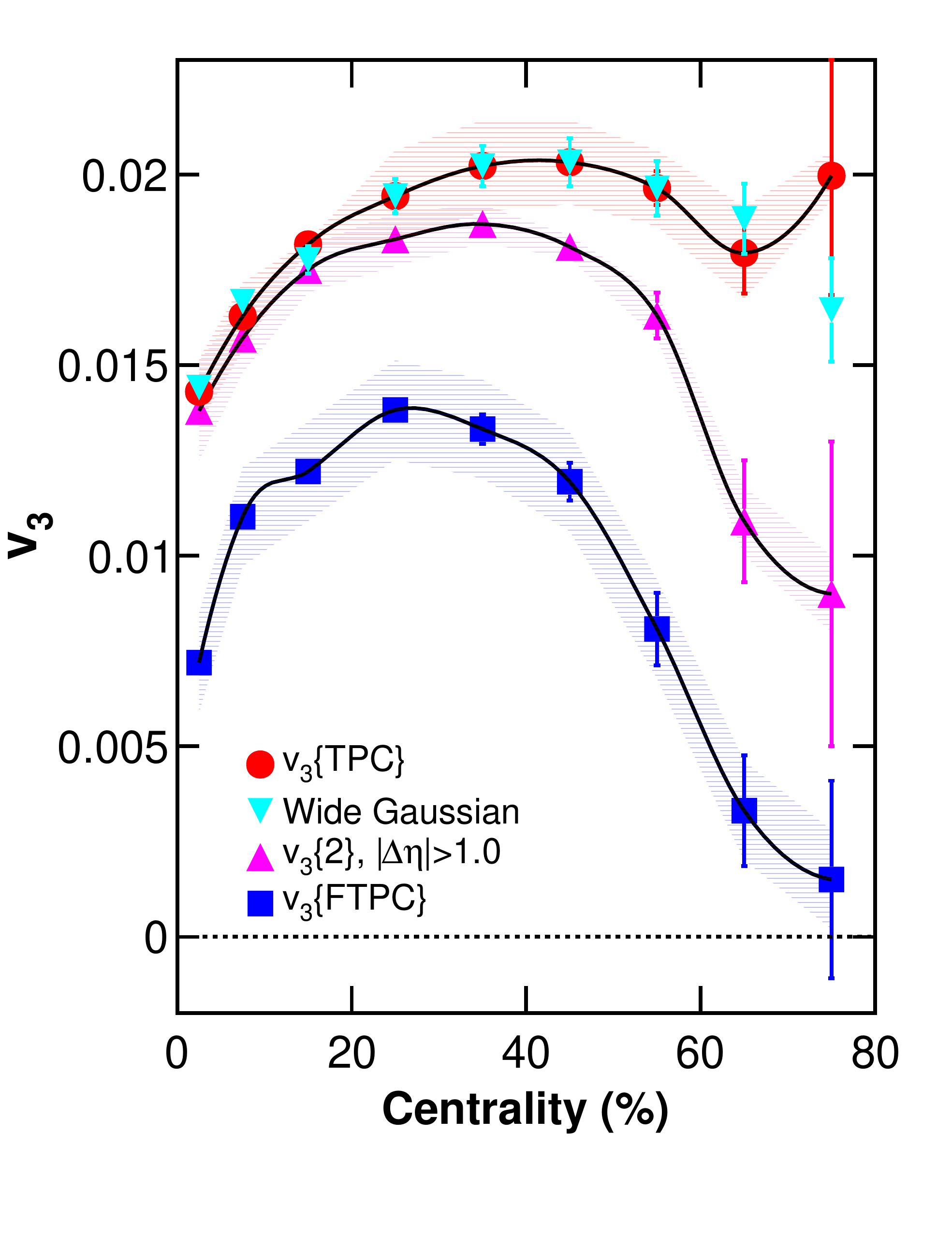} }
\caption {(color online) The third harmonic coefficient as a function of centrality from different methods of analysis for Au+Au collisions at $\sqrtsNN = 200$ GeV, integrated for $0.15 \lt p_T \lt 2.0$ \GeVc\ and $-1.0 \lt \eta \lt 1.0$. The curves connect the points and the bands show the systematic uncertainties. The systematic errors of the wide Gaussian method are similar to those for the TPC event plane method.}
\label{fig:cent}      
\end{figure}

Figure~\ref{fig:cent} shows the centrality dependence of $v_3$ obtained by integrating over $p_{T}$ using the observed yields. Shown are two-particle cumulants $v_{3}\{2\}$ with a minimum pseudorapidity separation between particles of one unit. Shown also is $v_{3}\{2\}$ from Eq.~(\ref{eq:int}) and Fig.~\ref{fig:gaus} for the wide Gaussian using particle pair weighting. Using weight $= 1$ in Eq.~(\ref{eq:int}) slightly lowered the wide Gaussian results for very peripheral collisions. Shown also are $v_{3}\{\rm{TPC}\}$ and $v_{3}\{\rm{FTPC}\}$ where $v_{3}$ is measured relative to the third harmonic event plane reconstructed either in the TPC subevents or the FTPCs. For $v_3\{2\}$ without a \deta cut the curve would be a factor of two higher for peripheral collisions and off scale.

Systematic uncertainties have been estimated by varying the DCA track cuts and the number of fit points, the event cut of vertex $z$, and the event plane flattening method. These uncertainties have been combined in quadrature to obtain the systematic uncertainties shown in Fig.~\ref{fig:cent}. The correlation of the third and second harmonic event planes was investigated by $\left< \cos 6 (\Psi_3-\Psi_2) \right>$ and within the statistical uncertainties was found to be consistent with zero for this data set. This is reasonable for this mixed harmonic result since observing the correlation between the third and second harmonic event planes requires a three particle correlation analysis to fix the direction of the first harmonic event plane [5].

\subsection{\deta dependence}
\label{sec:deltaEta}

\begin{figure*}
\resizebox{0.7\textwidth}{!}{
  \includegraphics{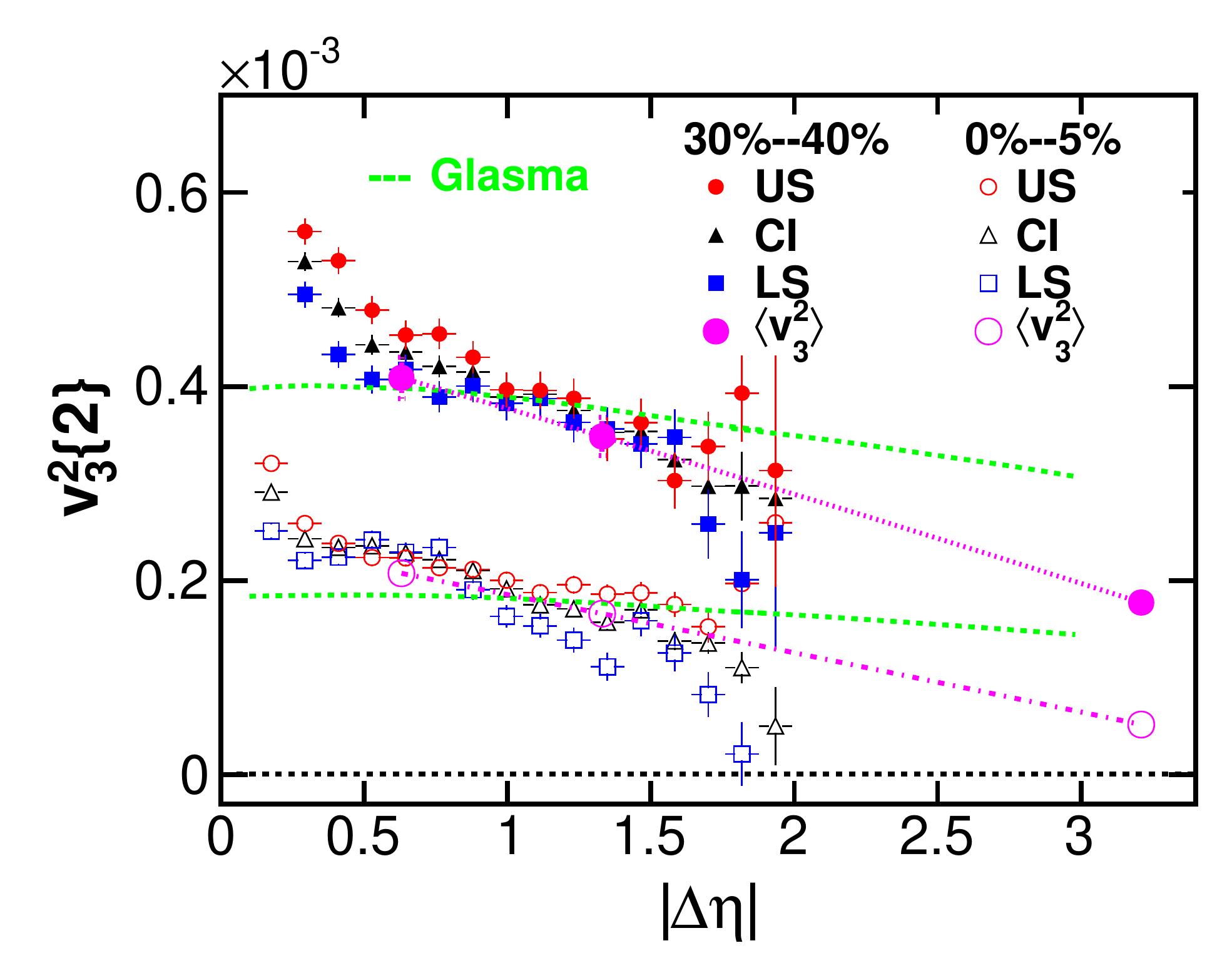} }
\caption {(color online) The square of the third harmonic coefficient as a function of pseudorapidity separation for Au+Au collisions at $\sqrtsNN = 200$ GeV for tracks with $0.15 \lt p_T \lt 2.0$ \GeVc. Shown are Unlike Sign (US), Charge Independent (CI), and Like Sign (LS) results at 0\%--5\% centrality (open symbols) and 30\%--40\% centrality (closed symbols). Most of the points at low $|\deleta|$ are not plotted because they correspond to the narrow Gaussian and go off the top of the scale. Also shown, by larger symbols, are the squares of the mean $v_3$ values (connected by purple dotted and dot-dashed lines) at the same two centralities from three analysis methods: The point at $|\deleta|=$ 0.63 is from the subevent method using the TPC with $ |\eta| \lt 1$. The point at 1.33 is from the 2-particle cumulant method with $|\Delta \eta| \gt 1$. The point at 3.21 is from correlations with the FTPC event plane. The dashed (green) curves without points are from a minimum bias Glasma calculation~\cite{DuslingPriv} for $\sqrtsNN = 200$ GeV with $0.15 \lt p_T \lt 2.0$ \GeVc\  done for the STAR acceptance with overall normalization set to the data at $|\deleta|$=1.} 
\label{fig:deltaEta}      
\end{figure*}

Clearly the various analysis methods for $v_{3}$ differ greatly in Fig.~\ref{fig:cent}. The results from the wide Gaussian and the TPC event plane are similar, showing that the narrow Gaussian effect is eliminated in both. When a large \deta is specified the $v_3$ values decrease, especially for the peripheral collisions in Fig.~\ref{fig:cent}. The  variation between most of the sets of results in Fig.~\ref{fig:cent} is caused by the \deta dependence as shown in Fig.~\ref{fig:deltaEta}. Two-particle correlation results in the TPC as a function of \deta for three charge combinations and two centralities are shown in Fig.~\ref{fig:deltaEta}. Also shown are the results for three analysis methods as a function of the mean \deta of the particles. For the points at $|\deleta|=$ 3.21 the event plane resolutions may be a bit low, and thus the $v_3$ values slightly high, because the $\eta$ gap between the two FTPCs is larger than that between the particles and the event plane. There is general agreement in the gradual decrease of $v_3$ with \deleta. The nonflow contributions due to short range correlations, seen as the narrow Gaussian in Fig.~\ref{fig:gaus}, are effectively suppressed by using either the wide Gaussian or by an $\eta$ gap. This result is consistent with previous studies of elliptic flow based on two-particle correlations, but in a previous work the corresponding wide Gaussian was ascribed to mini-jet correlations~\cite{minijet}. The decrease with \deta of $v_n^2\{2\}$ has been seen previously by the ATLAS Collaboration~\cite{ATLAS:2012at}. It has been calculated in Ref.~\cite{Bozek:2012en} as a decrease in nonflow. The decreasing effect of fluctuations from initial state gluon correlations has been described in Ref.~\cite{Dusling} but without evolution to the final state. The dilution of fluctuations during transport to the final state has been calculated in Ref.~\cite{Petersen:2011fp}. Reference~\cite{Xiao:2012uw} also describes the decorrelation of flow with increasing pseudorapidity gap using the AMPT model. Figure~\ref{fig:deltaEta} also is reminiscent of the well known near-side ridge in a plot of \deta vs.\ $\Delta \phi$ having a peak and shoulder~\cite{Voloshin:2011mx}. The far-side ridge may also contribute to this shoulder.

As Fig.~\ref{fig:deltaEta} shows, we did not find that $v_3$ stabilized at a constant value for large \deta within the acceptance of STAR. Thus one might ask if one should extrapolate to large \deta to avoid nonflow, or small \deta to measure all the fluctuations. However, it is clear that one must always quote \deta for each $v_3$ measurement and one must compare results to models with approximately the same \deta as the experiment. To help clarify the physics we compared like and unlike charge-sign combinations, because they have different contributions from resonance decays, fluctuations, and final state interactions, but we observed little difference between the combinations.  One source of fluctuations is calculated in the Glasma model~\cite{DuslingPriv} and shown by the Glasma lines, normalized to fit the data at \deta=1 in the figure. They show some decrease with \deleta, but not as much as in the data.

\subsection{Four-particle cumulants}

\begin{figure}
\resizebox{0.46\textwidth}{!}{
  \includegraphics{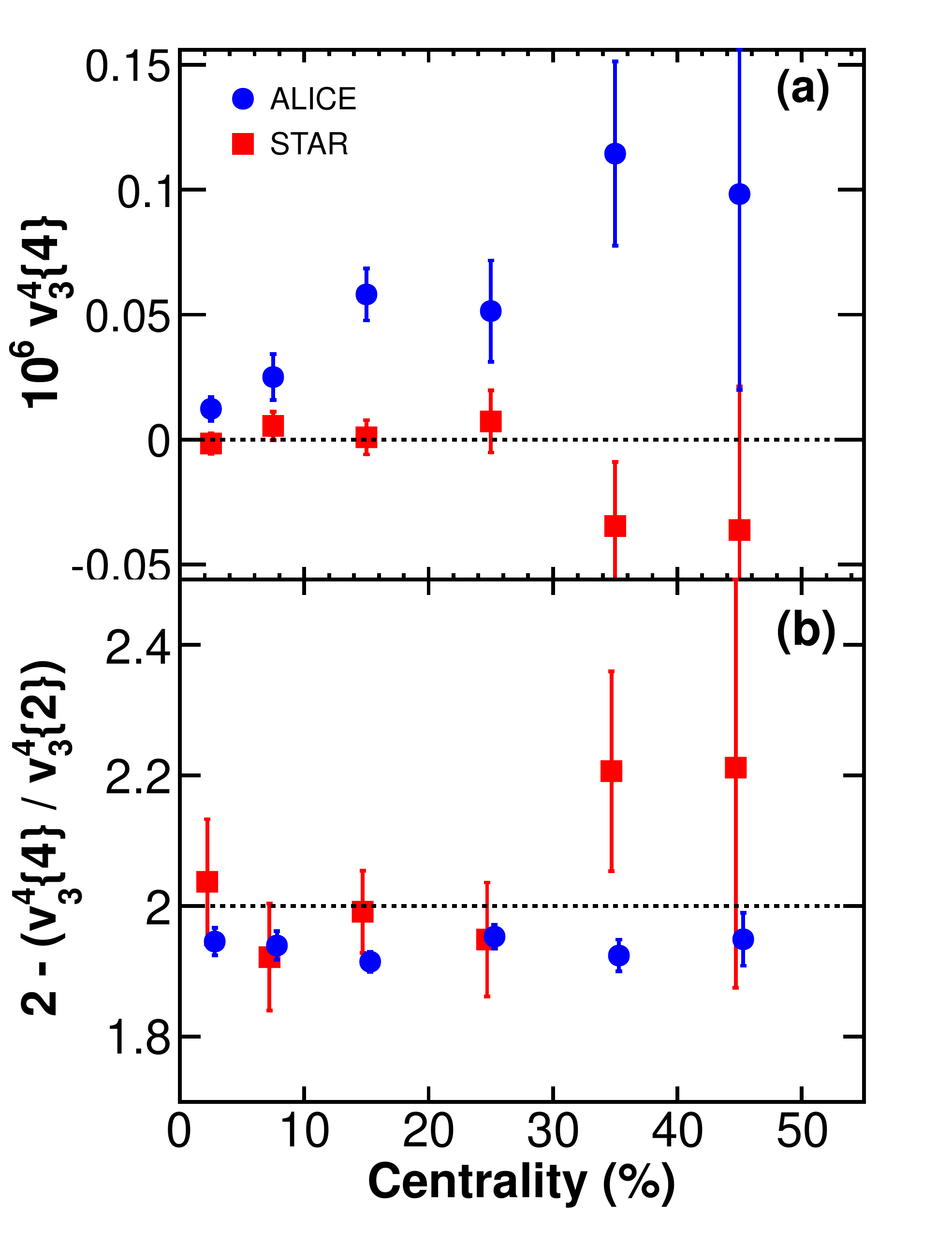} }
\caption {(color online) (a) The fourth power of the third harmonic coefficient from four-particle cumulants is plotted as a function of centrality for Au+Au collisions at $\sqrtsNN = 200$ GeV, with track selections $0.15 \lt p_T \lt 2.0$ \GeVc \ and $-1.0 \lt \eta \lt 1.0$. The ALICE results~\cite{ALICE:2011ab} are for Pb+Pb collisions at $\sqrtsNN = 2.76$ TeV, with track selections $0.2 \lt p_T \lt 5.0$ \GeVc \ and $-0.8 \lt \eta \lt 0.8$. (b) The points in the top figure are divided by the fourth power of the third harmonic flow from the $\eta$ subevent method, showing the deviation from 2.}
\label{fig:v34}      
\end{figure}

The results from four-particle cumulants, $v_{3}\{4\}$, with weighting by the number of combinations are shown in Fig.~\ref{fig:v34}~(a). They are consistent with zero within the errors, in contrast to the ALICE results~\cite{ALICE:2011ab} at the higher beam energy. Four-particle cumulants are known to suppress nonflow and Gaussian fluctuations~\cite{Sorensen:2011fb,Voloshin:2007pc}. To look for non-Gaussian fluctuations, Ref.~\cite{Bhalerao:2011ry} suggests plotting $(2*v_{3}^{4}\{2\} - v^{4}_{3}\{4\}) / v^{4}_{3}\{2\} = 2 - v^{4}_{3}\{4\} / v^{4}_{3}\{2\}$. This ratio, which is shown in Fig.~\ref{fig:v34}~(b), on the average overlaps with both the ALICE results and the expected Gaussian value of 2. Even though the differential $v_3(p_T)$ values for STAR and ALICE (which will be shown later) are the same, the integrated results for ALICE are larger, making their error bars in this figure smaller. Also, ALICE results come from a higher multiplicity at their higher beam energy, probably making the non-Gaussian effect more visible. Alternatively, the non-Gaussian fluctuations only may appear at the higher $p_T$ values included in the ALICE results. However, the precision of the STAR data does not allow us to conclude whether the STAR fluctuations are Gaussian or not.

\section{Comparisons to other experiments}

\begin{figure*}
\resizebox{0.54\textwidth}{!}{
  \includegraphics{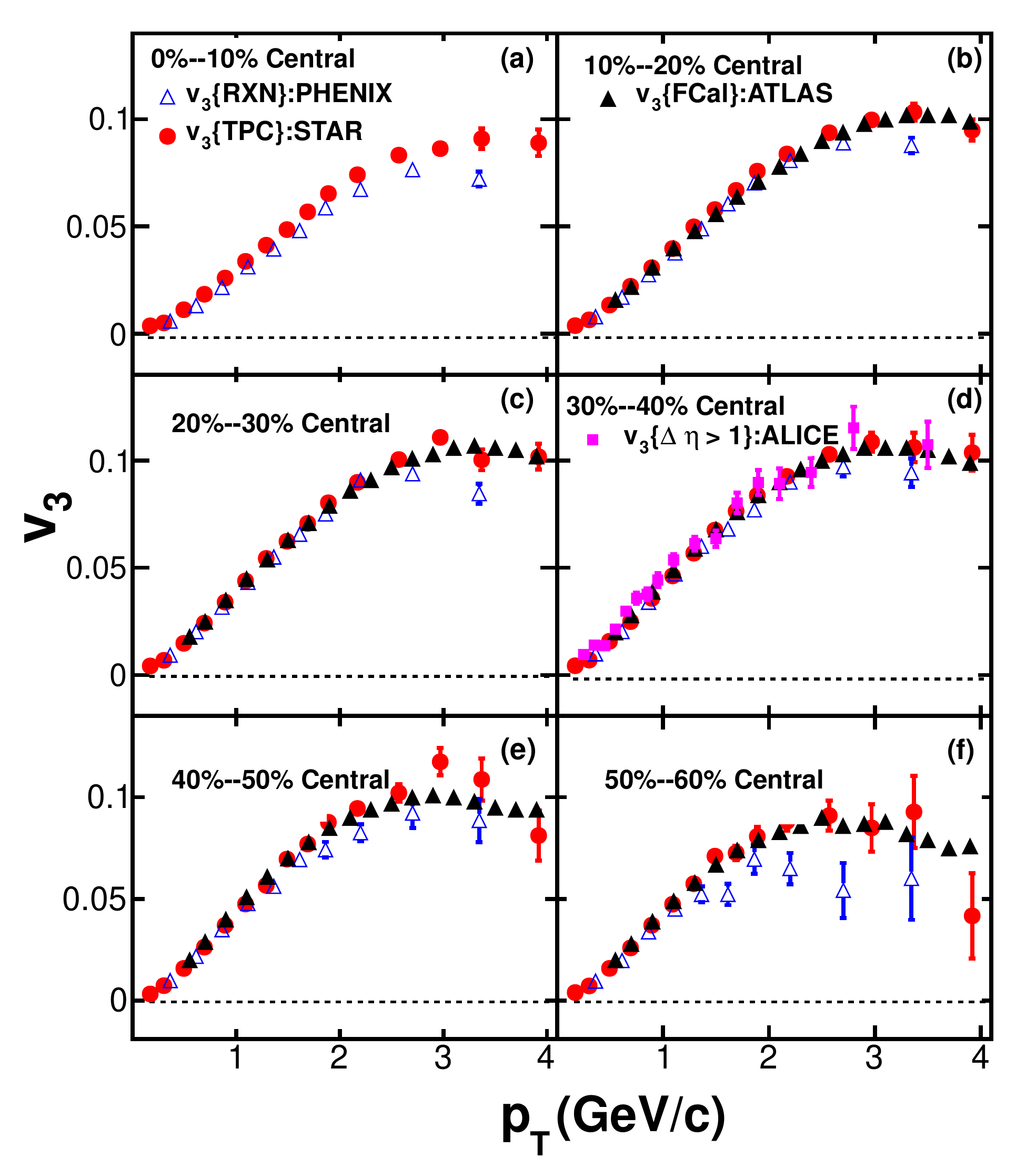} }
\caption {(color online) The third harmonic coefficient is plotted as a function of transverse momentum, for different centralities. The STAR $v_3\{\rm TPC\}$ results are from Fig.~\ref{fig:FTPC}. Also shown are PHENIX results, ATLAS results starting at 10\% centrality, and ALICE results for 30\%--40\% centrality.} 
\label{fig:LHC}      
\end{figure*}

Figure~\ref{fig:LHC} compares our $v_3\{\rm TPC\}$ results from Fig.~\ref{fig:FTPC} with those from PHENIX~\cite{Adare:2011tg}. 
The PHENIX results are shown for $|\eta| \le 0.35$, while for STAR the $\eta$ acceptance was $|\eta| \le 1.0$. For the STAR results from the TPC the mean $|\deleta|$ was 0.63, while for the results using the FTPC event plane the average $|\deleta|$ was 3.21. The PHENIX results used the event plane from their \rm{RXN} detector at an intermediate $\eta$ of $1.0 \lt \eta \lt 2.8$. Our results with the event plane in the TPC are very similar to those of PHENIX. This is surprising because the mean $\eta$ of their \rm{RXN} detector is larger than that for the subevents in our TPC. Our FTPC results in Fig.~\ref{fig:FTPC}, however, are lower than theirs. This is reasonable because the mean $|\deleta|$ is considerably larger in the FTPC than in the \rm{RXN} detector.

Comparison to LHC results for Pb+Pb at $\sqrtsNN$ = 2.76 TeV for ALICE~\cite{ALICE:2011ab} and ATLAS~\cite{ATLAS:2012at} are also shown in Fig.~\ref{fig:LHC}. ALICE results are for $|\eta| \lt 0.8$ and $|\deleta| \gt 1.0$. ATLAS results are for $|\eta| \lt 2.5$ with the event plane in the forward calorimeter at $3.2 \lt \eta \lt 4.9$, giving $|\deleta| \gt 0.8$. Agreement is good not only between RHIC experiments, but also between RHIC and LHC experiments. This is surprising because of the somewhat different \deta ranges.

\section{Model Comparisons}

\begin{figure*}
\resizebox{0.72\textwidth}{!}{
  \includegraphics{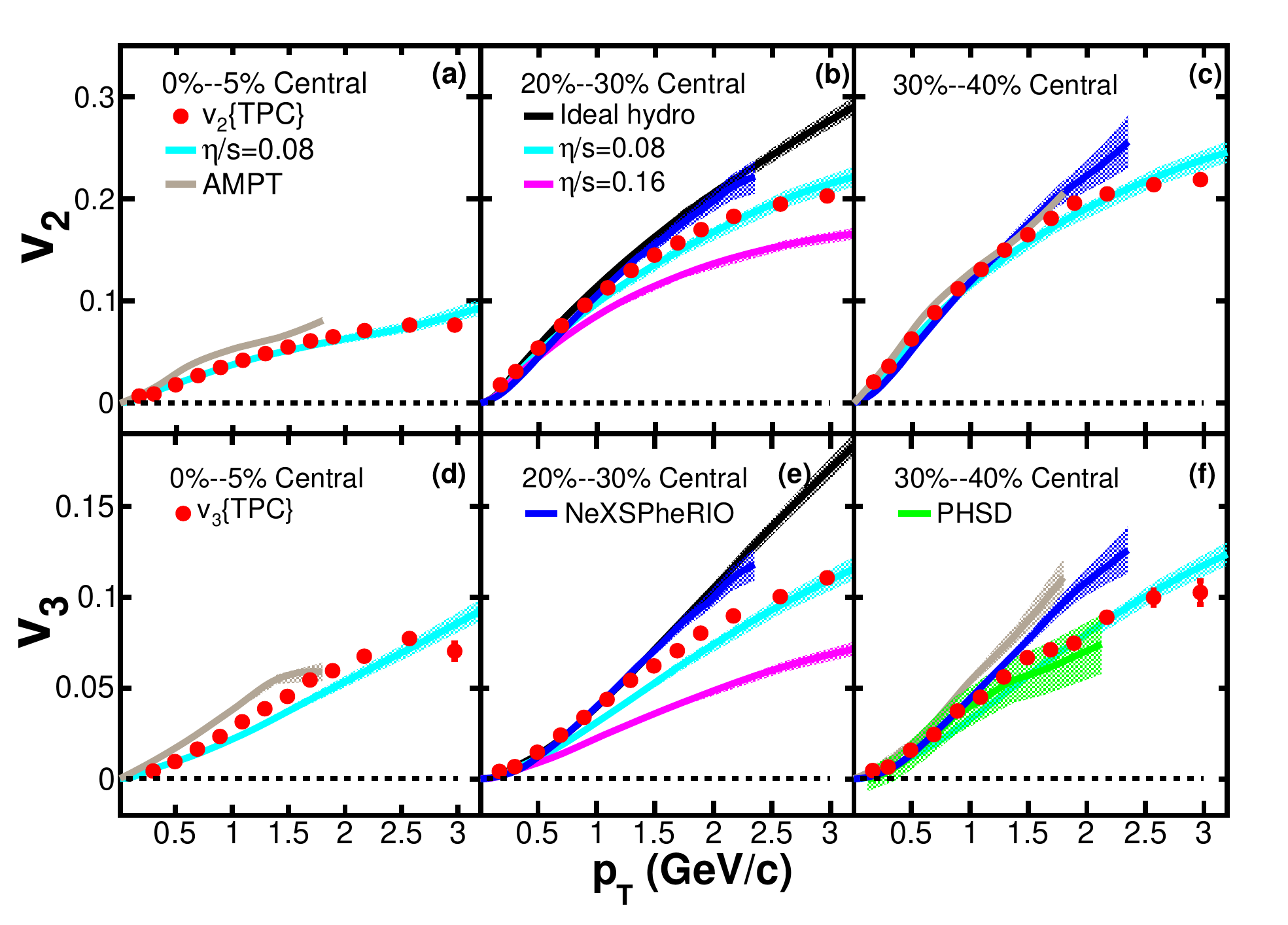}}
\caption{(color online) $v_2$ (top) and $v_3$ (bottom) for Au+Au collisions at $\sqrtsNN = 200$ GeV in 0\%--5\% (left), 20\%--30\% (middle), and 30\%--40\% (right) centrality as a function of transverse momentum at midrapidity, compared with ideal~\cite{Schenke:2011bn} (b),(e) and viscous hydro~\cite{Schenke:2011bn} (all), AMPT transport~\cite{v3-AMPT} (a),(c),(d),(f), NeXSPheRIO~\cite{Gardim:2012yp} (b),(c),(e),(f), and Parton Hardon String Dynamics~\cite{Konchakovski:2012yg} (f) models. The STAR $v_2$ values (top) are from Ref.~\cite{Adams:2004bi}.}
\label{fig:models} 
\end{figure*}

In the event-by-event ideal hydro model, $v_3$ was studied first by Ref.~\cite{v3_4D-hydro}, and then by Ref.~\cite{Gardim:2012yp}. References~\cite{Qiu:2011iv,Qiu:2011hf} concluded that instead of averaged initial conditions, event-by-event calculations are necessary to compare with experimental data. The first prediction of $v_3$ with viscous hydro was in Ref.~\cite{transpov3}. Recent reviews of viscous hydro have been presented in Refs.~\cite{Heinz:2013th,Gale:2013da}. The linear translation from initial space fluctuations to final momentum fluctuations has been calculated for elliptic flow with the NeXSPheRIO model~\cite{deSouza:2011rp}. Reference~\cite{Kapusta:2011gt} calculates the additional fluctuations induced during the viscous expansion. 

\subsection{Pseudorapidity separation}
Calculations of $v_3^2\{2\}$ vs.\ \deta have been done in Ref.~\cite{Bozek:2012en}. They used an event-by-event viscous hydro model and addressed the effect of radial flow on local charge conservation in hadronization. Their results have a similar $v_3^2\{2\}$ vs.\ \deta slope as the data in Fig.~\ref{fig:deltaEta}, but the values are higher than the data. The normalization to fit the data probably could be adjusted. But their charge balancing mechanism would predict a much bigger difference between unlike-sign pairs and like-sign pairs. There is only a small spread in the data in Fig.~\ref{fig:deltaEta} at \deta about 0.5, largely ruling out this mechanism. 

The Glasma model calculations of Ref.~\cite{DuslingPriv} show some decrease in $v_3^2\{2\}$ with $|\deleta|$ in Fig.~\ref{fig:deltaEta} giving a partial explanation for the decrease with $|\deleta|$. However, these calculations for the initial state are not sufficient to explain the sharper fall off of $v_3^2\{2\}$ vs.\ $|\deleta|$ seen in the data. This perturbative model is strictly only valid at the higher $p_T$ values ($p_T \gg Q_S$, where $Q_S$ is the saturation scale of the bulk matter produced in the collision). Reference~\cite{DuslingPriv} says ``The decorrelation of the two-particle correlation with increasing rapidity gap demonstrates the violation of the boost invariance of the classical Glasma flux tube picture by quantum evolution effects." In principle the normalization could be determined by hydrodynamic transport to the final state. However, it is probable that the large discrepancy between the methods in Fig.~\ref{fig:cent} has its origin in the \deta dependence of fluctuations, either in the initial state or in the hydrodynamic evolution.

Another Glasma flux tube model with radial flow has been used to calculate fluctuations and $v_3$~\cite{Gavin:2012if}. Reference~\cite{Voloshin:2011mx} says that the near-side ridge caused by long-range $\eta$ correlations, and odd harmonics in the azimuthal anisotropy, are two ways of describing the same phenomenon, \emph{i.e.} the response of the system to fluctuations in the initial density distribution.

\subsection{Transverse momentum dependence}
In Fig.~\ref{fig:models}, $v_2$~\cite{Adams:2004bi} and $v_3$ obtained with the TPC subevent plane method are compared as a function of transverse momentum with several models for 0\%--5\%, 20\%--30\%, and 30\%--40\% central collisions. The experimental results for the TPC subevent plane method are shown because they eliminate the short-range correlations but yet have a small $|\deleta|$ like the theory calculations. Shown in Fig.~\ref{fig:models} are the ideal and the viscous event-by-event hydrodynamic model of Refs.~\cite{Schenke:2011bn,Gale:2012rq} where the initial conditions come from a Monte Carlo Glauber model and the ratio of shear viscosity ($\eta$) to entropy density ($s$) is $\eta/s = $ 0.0 (ideal), 0.08, and 0.16. To properly include fluctuations, 100 to 200 events were  simulated and then the root-mean-square flow values calculated. The agreement with the hydro for $\eta/s = 0.08$ is very good. NeXSPheRIO~\cite{Gardim:2012yp} root-mean-square results for 20\%--30\% and 30\%--40\%  centralities at $p_T$ below one \GeVc \ are also good. Also shown are the results from the AMPT model~\cite{v3-AMPT} with string melting for the latest set of parameters (``Set B''). The agreement for $v_2$ is good, but the calculated $v_3$ is a bit high in panels (d) and (f). AMPT has also been used for $v_3$ from symmetric~\cite{Xiao:2011ti, Solanki:2012ne} and asymmetric collisions~\cite{Haque:2011ti}. Predictions for $v_3$ from Parton Hadron String Dynamics~\cite{Konchakovski:2012yg} at 30\%--40\% centrality for $|\eta| \lt 0.5$ have been made by the subevent method with the event planes at $1.0 \lt |\eta| \lt 4.0$, and show good agreement in the figure lower right. HIJING~\cite{HIJING} does not predict any significant $v_3$ as $v_3^2$ in the $p_T$ range up to 1.5 GeV/$c$ is both negative and positive, with absolute values less than $2 \times 10^{-4}$, and is therefore not shown in Fig.~\ref{fig:models}.

Elliptic flow results have been mostly described by hydro with $\eta/s = $ 0.08 with Glauber initial conditions in the case of midcentral collisions~\cite{Schenke:2011bn}. We find that the $v_3$ results are described by this model with a similar viscosity. The NeXSPheRIO model at low $p_T$ and the PHSD model also agree with the data.

\section{Summary}
We have presented measurements of third harmonic flow of charged particles from Au+Au collisions at $\sqrtsNN = 200$ GeV as a function of pseudorapidity, transverse momentum, pseudorapidity gap, charge sign, and centrality made with the STAR detector at RHIC. We have reported results from a two-particle method for particle pairs with an $\eta$ gap or fit with a wide Gaussian in pseudorapidity separation, as well as from the standard event-plane method with the event plane near midrapidity or at forward rapidity. Short-range correlations are eliminated either by an $\eta$ gap or by discarding the narrow Gaussian in pseudorapidity separation. The measured values of $v_3$ continuously decrease as the mean pseudorapidity separation of the particles increases within the range observable by STAR. A model for nonflow predicts a big difference between different charge sign pairs which is not observed in the data. A model for the decrease of fluctuations with pseudorapidity separation from a Glasma~\cite{DuslingPriv} initial state reproduces some aspects of the data.  Because of this, and the good agreement of $v_3(p_T)$ with models including fluctuations, it is likely that $v_3$ is mainly due to \deta dependent fluctuations~\cite{Dusling}. According to the models, these fluctuations should be largely independent of beam energy. 


\begin{acknowledgments}
For supplying data and model calculations we thank Ante Bilandzic (ALICE), Kevin Dusling (Glasma model), Fernando Gardim (NeXSPheRIO), Jiangyong Jia (ATLAS), Che-Ming Ko (AMPT), Volodya Konchakovski (PHSD), Bjoern Schenke (hydro), Raimond Snellings (ALICE), and Jun Xu (AMPT). We also benefited greatly from conversations with Jean-Yves Ollitrault, Rajeev Bhalerao, and Kevin Dusling.

We thank the RHIC Operations Group and RCF at BNL, the NERSC Center at LBNL and the Open Science Grid consortium for providing resources and support. This work was supported in part by the Offices of NP and HEP within the U.S. DOE Office of Science, the U.S. NSF, the Sloan Foundation, CNRS/IN2P3, FAPESP CNPq of Brazil, Ministry of Ed. and Sci. of the Russian Federation, NNSFC, CAS, MoST, and MoE of China, GA and MSMT of the Czech Republic, FOM and NWO of the Netherlands, DAE, DST, and CSIR of India, Polish Ministry of Sci. and Higher Ed., National Research Foundation (NRF-2012004024), Ministry of Sci., Ed. and Sports of the Rep. of Croatia, and RosAtom of Russia.

\end{acknowledgments}


\normalsize
\begin{thebibliography}{99}

\bibitem{Voloshin:2008dg}
  S.~A.~Voloshin, A.~M.~Poskanzer and R.~Snellings,
  in Landolt-Boernstein, {\em Relativistic Heavy Ion Physics}, Vol. 1/23, p. 5-54 (Springer-Verlag, 2010),
  arXiv:0809.2949 [nucl-ex].  
  
\bibitem{Sorensen:2009cz} 
  P.~Sorensen,
  arXiv:0905.0174 [nucl-ex];
  In Quark-Gluon Plasma 4 by R. Hwa and X.N. Wang, World Scientific (2010).
  
\bibitem{v4v6}
  P. F.~Kolb, Phys. Rev. C {\bf 68}, 031902(R) (2003);
  J.~Adams {\it et al.} [STAR Collaboration], Phys. Rev. Lett. {\bf 92}, 062301 (2004).

\bibitem{Mishra:2007tw}
 A.~P.~Mishra, R.~K.~Mohapatra, P.~S.~Saumia and A.~M.~Srivastava,
 Phys.\ Rev.\ C {\bf 77}, 064902 (2008),
 [arXiv:0711.1323 [hep-ph]]

\bibitem{geoFluct1}
  B.~Alver and G.~Roland,
  Phys.\ Rev.\ C {\bf 81}, 054905 (2010)
  [Erratum ibid.\ C {\bf 82}, 039903 (2010)],
  [arXiv:1003.0194 [nucl-th]].
  
\bibitem{derik} 
  D.~Teaney and L.~Yan,
  Phys.\ Rev.\ C {\bf 83}, 064904 (2011),
  [arXiv:1010.1876 [nucl-th]].

\bibitem{Sorensen:2011hm} 
  P.~Sorensen, B.~Bolliet, A.~Mocsy, Y.~Pandit and N.~Pruthi,
  Phys.\ Lett.\ B {\bf 705}, 71 (2011),
  [arXiv:1102.1403 [nucl-th]].
 
\bibitem{Ollitrault:2009ie} 
  J.~-Y.~Ollitrault, A.~M.~Poskanzer and S.~A.~Voloshin,
  Phys.\ Rev.\ C {\bf 80}, 014904 (2009),
  [arXiv:0904.2315 [nucl-ex]].

\bibitem{Mishra}
  A.~P.~Mishra, R.~K.~Mohapatra, P.~S.~Saumia and A.~M.~Srivastava,
  Phys.\ Rev.\ C {\bf 81}, 034903 (2010),
  [arXiv:0811.0292 [hep-ph]].
  
\bibitem{geoFluct2}
  G.~-Y.~Qin, H.~Petersen, S.~A.~Bass and B.~Muller,
  Phys.\ Rev.\ C {\bf 82}, 064903 (2010),
  [arXiv:1009.1847 [nucl-th]].

\bibitem{transpov3}
  B.~H.~Alver, C.~Gombeaud, M.~Luzum and J.~-Y.~Ollitrault,
  Phys.\ Rev.\ C {\bf 82}, 034913 (2010),
  [arXiv:1007.5469 [nucl-th]].
  
\bibitem{v3_4D-hydro}
  H.~Petersen, G.~-Y.~Qin, S.~A.~Bass and B.~Muller,
  Phys.\ Rev.\ C {\bf 82}, 041901 (2010),
  [arXiv:1008.0625 [nucl-th]].
  
\bibitem{hydrov3}
  B.~Schenke, S.~Jeon and C.~Gale,
  Phys.\ Rev.\ Lett.\  {\bf 106}, 042301 (2011),
  [arXiv:1009.3244 [hep-ph]].
  
\bibitem{Schenke:2011bn} 
  B.~Schenke, S.~Jeon and C.~Gale,
  Phys.\ Rev.\ C {\bf 85}, 024901 (2012),
  [arXiv:1109.6289 [hep-ph]].
  
\bibitem{v3-AMPT} 
  J.~Xu and C.~M.~Ko,
  Phys.\ Rev.\ C {\bf 84}, 014903 (2011),
  [arXiv:1103.5187 [nucl-th]].

\bibitem{Schenke:2012wb} 
  B.~Schenke, P.~Tribedy and R.~Venugopalan,
  Phys.\ Rev.\ Lett.\  {\bf 108}, 252301 (2012),
  [arXiv:1202.6646 [nucl-th]].
  
\bibitem{Gale:2012rq} 
  C.~Gale, S.~Jeon, B.~Schenke, P.~Tribedy and R.~Venugopalan,
  Phys.\ Rev.\ Lett.\  {\bf 110}, 012302 (2013),
  arXiv:1209.6330 [nucl-th].

\bibitem{Voloshin:2011mx} 
  S.~A.~Voloshin,
  Prog.\ Part.\ Nucl.\ Phys.\  {\bf 67}, 541 (2012),
  [arXiv:1111.7241 [nucl-ex]].
  
\bibitem{Bhalerao:2011yg} 
  R.~S.~Bhalerao, M.~Luzum and J.~-Y.~Ollitrault,
  Phys.\ Rev.\ C {\bf 84}, 034910 (2011),
  [arXiv:1104.4740 [nucl-th]].
  
\bibitem{Luzum:2010fb} 
  M.~Luzum and J.~-Y.~Ollitrault,
  Phys.\ Rev.\ Lett.\  {\bf 106}, 102301 (2011),
  [arXiv:1011.6361 [nucl-ex]].

\bibitem{Teaney:2012ke} 
  D.~Teaney and L.~Yan,
  Phys.\ Rev.\ C {\bf 86}, 044908 (2012),
  [arXiv:1206.1905 [nucl-th]].
  
\bibitem{Cumulant} 
  A.~Bilandzic, R.~Snellings and S.~Voloshin,
  Phys.\ Rev.\ C {\bf 83}, 044913 (2011),
  [arXiv:1010.0233 [nucl-ex]].

\bibitem{methodPaper}
A. M. Poskanzer and S. A. Voloshin, Phys. Rev. C {\bf 58}, 1671 (1998).

\bibitem{TPC}
	M.~Anderson \etal, Nucl. Instrum. Meth. A {\bf 499}, 659 (2003).

\bibitem{FTPC}
	K.~H.~Ackermann \etal, Nucl. Instrum. Meth. A {\bf 499}, 713 (2003).
	
\bibitem{Agakishiev:2011eq} 
  G.~Agakishiev {\it et al.}  [STAR Collaboration],
  Phys.\ Rev.\ C {\bf 86}, 014904 (2012),
  [arXiv:1111.5637 [nucl-ex]].

\bibitem{Alver:2010rt} 
  B.~Alver {\it et al.}  [PHOBOS Collaboration],
  Phys.\ Rev.\ C {\bf 81}, 034915 (2010),
  [arXiv:1002.0534 [nucl-ex]].

\bibitem{Kapusta:2011gt} 
  J.~I.~Kapusta, B.~Muller and M.~Stephanov,
  Phys.\ Rev.\ C {\bf 85}, 054906 (2012),
  [arXiv:1112.6405 [nucl-th]].
  
\bibitem{Dusling}
  K.~Dusling, F.~Gelis, T.~Lappi and R.~Venugopalan,
  Nucl.\ Phys.\ A {\bf 836}, 159 (2010),
  [arXiv:0911.2720 [hep-ph]].
  
\bibitem{Agakishiev:2011pe} 
  G.~Agakishiev {\it et al.}  [STAR Collaboration],
  Phys.\ Rev.\ C {\bf 86}, 064902 (2012),
  arXiv:1109.4380 [nucl-ex].

\bibitem{minijet}
   M. Daugherity [STAR Collaboration], J. Phys. G {\bf 35}, 104090 (2008); D. Kettler [STAR Collaboration], PoS C ERP2010, 011 (2010); T. A. Trainor and D. T. Kettler, Phys. Rev. C {\bf 83}, 034903 (2011);
   T.~A.~Trainor, D.~J.~Prindle and R.~L.~Ray,
  Phys.\ Rev.\ C {\bf 86}, 064905 (2012),
  [arXiv:1206.5428 [hep-ph]].

\bibitem{ATLAS:2012at} 
  G.~Aad {\it et al.}  [ATLAS Collaboration],
  Phys.\ Rev.\ C {\bf 86} 014907 (2012), 
  [arXiv:1203.3087 [hep-ex]].

\bibitem{Bozek:2012en} 
  P.~Bozek and W.~Broniowski,
  Phys.\ Rev.\ Lett.\  {\bf 109}, 062301 (2012),
  [arXiv:1204.3580 [nucl-th]].
  
\bibitem{Petersen:2011fp} 
  H.~Petersen, V.~Bhattacharya, S.~A.~Bass and C.~Greiner,
  Phys.\ Rev.\ C {\bf 84}, 054908 (2011),
  [arXiv:1105.0340 [nucl-th]].

\bibitem{Xiao:2012uw} 
  K.~Xiao, F.~Liu and F.~Wang,
  Phys.\ Rev.\ C {\bf 87}, 011901 (2013),
  [arXiv:1208.1195 [nucl-th]].
  
\bibitem{DuslingPriv}
 K. Dusling, private communication (2012).

\bibitem{ALICE:2011ab} 
  K.~Aamodt {\it et al.}  [ALICE Collaboration],
  Phys.\ Rev.\ Lett.\  {\bf 107}, 032301 (2011),
  [arXiv:1105.3865 [nucl-ex]].
  
\bibitem{Sorensen:2011fb} 
  P.~Sorensen [STAR Collaboration],
  J.\ Phys.\ G {\bf 38}, 124029 (2011),
  [arXiv:1110.0737 [nucl-ex]].

\bibitem{Voloshin:2007pc} 
  S.~A.~Voloshin, A.~M.~Poskanzer, A.~Tang and G.~Wang,
  Phys.\ Lett.\ B {\bf 659}, 537 (2008),
  [arXiv:0708.0800 [nucl-th]].

\bibitem{Bhalerao:2011ry} 
  R.~S.~Bhalerao, M.~Luzum and J.~Y.~Ollitrault,
  J.\ Phys.\ G {\bf 38}, 124055 (2011),
  [arXiv:1106.4940 [nucl-ex]].
    
\bibitem{Adare:2011tg} 
  A.~Adare {\it et al.}  [PHENIX Collaboration],
  Phys.\ Rev.\ Lett.\  {\bf 107}, 252301 (2011),
  [arXiv:1105.3928 [nucl-ex]].
  
\bibitem{Gardim:2012yp} 
  F.~G.~Gardim, F.~Grassi, M.~Luzum and J.~-Y.~Ollitrault,
  Phys.\ Rev.\ Lett.\  {\bf 109}, 202302 (2012),
  [arXiv:1203.2882 [nucl-th]].
  
\bibitem{Qiu:2011iv} 
  Z.~Qiu and U.~W.~Heinz,
  Phys.\ Rev.\ C {\bf 84}, 024911 (2011),
  [arXiv:1104.0650 [nucl-th]].
  
\bibitem{Qiu:2011hf} 
  Z.~Qiu, C.~Shen and U.~Heinz,
  Phys.\ Lett.\ B {\bf 707}, 151 (2012),
  [arXiv:1110.3033 [nucl-th]].
  
\bibitem{Heinz:2013th} 
  U.~Heinz and R.~Snellings,
  arXiv:1301.2826 [nucl-th].

\bibitem{Gale:2013da} 
  C.~Gale, S.~Jeon, and B.~Schenke,
  Int.\  J.\  of Mod.\  Phys.\  A, Vol.\  28, {\bf 1340011} (2013),
  [arXiv:1301.5893 [nucl-th]].
  
\bibitem{deSouza:2011rp} 
  R.~D.~de Souza, J.~Takahashi, T.~Kodama and P.~Sorensen,
  Phys.\ Rev.\ C {\bf 85}, 054909 (2012),
  [arXiv:1110.5698 [hep-ph]].
  
\bibitem{Gavin:2012if} 
  S.~Gavin and G.~Moschelli,
  Phys.\ Rev.\ C {\bf 86}, 034902 (2012),
  [arXiv:1205.1218 [nucl-th]].
  
\bibitem{Adams:2004bi} 
  J.~Adams {\it et al.}  [STAR Collaboration],
  Phys.\ Rev.\ C {\bf 72}, 014904 (2005),
  [nucl-ex/0409033].
  
\bibitem{Xiao:2011ti} 
  K.~Xiao, N.~Li, S.~Shi and F.~Liu,
  J.\ Phys.\ G {\bf 39}, 025011 (2012),
  [arXiv:1111.6213 [nucl-th]].

\bibitem{Solanki:2012ne} 
  D.~Solanki, P.~Sorensen, S.~Basu, R.~Raniwala and T.~K.~Nayak,
  arXiv:1210.0512 [nucl-ex].
  
\bibitem{Haque:2011ti} 
  M.~R.~Haque, M.~Nasim and B.~Mohanty,
  Phys.\ Rev.\ C {\bf 84}, 067901 (2011),
  [arXiv:1111.5095 [nucl-ex]].
  
\bibitem{Konchakovski:2012yg} 
  V.~P.~Konchakovski, E.~L.~Bratkovskaya, W.~Cassing, V.~D.~Toneev, S.~A.~Voloshin and V.~Voronyuk,
  Phys.\  Rev.\  {\bf C85}, 044922 (2012),
  [arXiv:1201.3320 [nucl-th]].

 \bibitem{HIJING}
 X.~N.~Wang and M.~Gyulassy, 
Phys.\ Rev.\ {\bf D 44}, 3501 (1991).

 \end{thebibliography}
\end{document}